PUC

# A Note on Process Modelling: Combining Situation Calculus and Petri Nets

**Edirlei Soares de Lima**
**Antonio L. Furtado**
**Bruno Feijó**
**Marco A. Casanova**

Departamento de Informática

**PONTIFÍCIA UNIVERSIDADE CATÓLICA DO RIO DE JANEIRO**
**RUA MARQUÊS DE SÃO VICENTE, 225 - CEP 22451-900**
**RIO DE JANEIRO - BRASIL**

# A Note on Process Modelling: Combining Situation Calculus with Petri Nets

Edirlei Soares de Lima, Antonio L. Furtado, Bruno Feijó, Marco A. Casanova

Universidade Europeia, Faculty of Design, Technology and Communication – IADE, Lisbon, Portugal
PUC-Rio, Departamento de Informática, Rio de Janeiro, Brasil
edirlei.lima@universidadeeuropeia.pt, {Furtado,Bruno,Casanova}@inf.puc-rio.br

**Abstract:** The *situation calculus logic model* is most convenient for modelling the actions that can occur in an information system application. The interplay of pre-conditions and post-conditions not only determines a semantically justified partial order of the defined actions, but also serves to enforce integrity constraints. This form of specification allows the use of plan-generation algorithms to investigate, before the system is liberated for official usage, whether the proposed specification allows all desirable use cases, and effectively disallows the illegal or, for some reason, undesirable ones. On the other hand, especially for legacy applications, implemented without a prior specification, Process Mining techniques have been employed to derive an implicit *Petri net model* from the analysis of a large enough number of traces registered in an execution log. However, as a system just begins to be used, with a still empty execution log, this sort of process mining discovery would not be feasible. We shall explain in this paper how the Petri net model can be directly derived from the situation calculus specification rules. The main gist of the present research is to provide evidence that the two models are complementary, not only because the Petri net model is derivable from the situation calculus model, but also in view of the distinct advantages of the two models. While the situation calculus model leads to planning and simulated execution prior to implementation, the Petri net model, like other workflow engines, can be designed to run in a tightly restrictive mode, with the additional asset of intuitive visualization of the workable sequences. As proof of concept, we developed a prototype to demonstrate our methods, and tried it on two example cases: 1. a published *request processing* application used to introduce process mining notions, and 2. an analogously structured *trial by combat* application taken from a popular movie. The prototype includes an interactive dramatization component, which serves – as a mandatory requirement for digital entertainment pieces – to enact the second application.

**Resumo:** O modelo lógico de *cálculo de situação* é conveniente para modelar as ações que podem ocorrer em uma aplicação de sistema de informação. O jogo de pré-condições e pós-condições não apenas determina uma ordem parcial semanticamente justificada, como também serve para garantir restrições de integridade. Essa forma de especificação permite o uso de algoritmos de geração de planos para investigar, antes de o sistema ser oficialmente liberado para uso, se a especificação proposta permite todos os casos de uso desejáveis, e efetivamente proíbe os casos ilegítimos ou, por alguma razão, considerados indesejáveis. Por outro lado, especialmente para aplicações herdadas, implementadas sem prévia especificação, técnicas de Mineração de Processos tem sido empregadas para derivar um modelo de rede de Petri implícito a partir da análise de uma quantidade suficiente de traços registrados em um log de execução. Contudo, no momento em que um sistema apenas começa a ser usado, com um log de execução ainda vazio, essa espécie de mineração de processos não seria factível. Vamos explicar neste artigo como o modelo de rede de Petri pode ser derivado diretamente das regras de especificação do cálculo de situação. O ponto essencial da presente pesquisa é fornecer prova de que os dois modelos são complementares, não apenas porque o modelo de rede de Petri é derivável do modelo de cálculo de situação, mas também em vista das vantagens distintas dos dois modelos. Enquanto o modelo de cálculo de situação leva a planejamento e a execução simulada antes da implementação, o modelo de rede de Petri, assim como outras ferramentas de fluxo de trabalho, pode ser projetado para rodar em um modo estritamente restritivo, com o bônus adicional da visualização intuitiva das sequências viáveis. Como prova de conceito, desenvolvemos um protótipo para demonstrar nossos métodos, e o experimentamos em dois casos exemplos: 1. uma aplicação de *processamento de pedidos* publicada para introduzir noções de mineração de processos, e 2. uma aplicação, de estrutura análoga, de *julgamento por combate* tirada de um filme popular. O protótipo inclui um componente de dramatização interativo, que serve – como requisito obrigatório em peças de entretenimento digital – para encenar a segunda aplicação.

Rosane Teles Lins Castilho
Assessoria de Biblioteca, Documentação e Informação
PUC-Rio Departamento de Informática
Rua Marquês de São Vicente, 225 - Gávea
22451-900 Rio de Janeiro RJ Brasil
Tel. +55 21 3527-1516 Fax: +55 21 3527-1530
E-mail: bib-di@inf.puc-rio.br

## 1. Introduction

It is highly recommended that the implementation of Information Systems – both for business applications and for digital entertainment – be preceded (and governed) by a conceptual specification. Typically, formalisms such as Entity-Relationship or UML models are employed to specify the database component. But the processes whereby the database is handled should also be specified.

Systems implemented without a conceptual specification (legacy systems, schema-less systems) are unpredictable and problematic. Reverse engineering techniques have been used to move up from arbitrarily designed files to a fitting conceptual specification [Heuser].

Another way to obtain a conceptual specification, when none is available and the system has been running for some time, is to apply *process mining* techniques [Aalst]. A process model can be obtained by extracting traces from an execution log, analyzing and somehow combining them, and finally expressing the result under some formal representation (e.g. as a *Petri net*).

Here we shall first employ our own formal approach, based on situation calculus. In previous works (e.g. [Gottin]) we have shown how to start with a formal conceptual specification and ultimately derive a relational database implementation by going through 3 phases: (1) three-schema executable specification in Prolog, (2) execution commanded from the Prolog environment resulting in the update of Oracle tables, (3) final Oracle implementation wherein actions are performed via automatically generated stored procedures.

As part of that previous work, we demonstrated an early version of process mining, by extracting transactions from an also automatically created execution log.

The situation calculus specification that we use as proof of concept in this report shows how we interpret the simple introductory Petri net example, formulated by Wil van der Aalst [Aalst], which deals with a reimbursement request process. Starting from the logic specification has the advantage of offering a semantic meaning to transactions, as well as putting goal-motivated plans side by side with observed traces.

We shall argue that the two models are complementary to each other, to the point that the Petri net representation can be generated from the situation calculus model, over which an execution method, analogous to the standard token-based Petri net method, can be operated.

In addition, as we have argued before [Ciarlini et al. 2010], our very same logical approach is applicable to the interactive composition of plots pertaining to fictional storytelling genres. As a support to this claim, we shall examine how a trial-by-combat incident taken from the movie *Excalibur*[1] can be handled in terms of the two models. The choice of this particular incident was largely motivated by the felicitous coincidence of its Petri net structure with that of the request application.

Related research includes in special [Tan], which proposes a Situation Calculus Ontology of Petri Nets. In [Peterson] a comprehensive and still useful classic survey of the Petri net formalism is provided. Petri net synthesis is preceded by a novel preliminary activity mining algorithm in [Kindler et al.]. Plans are related to Petri nets by [Ziparo et al.] and plans are used to produce workflows in [Fernandes et. al] (recalling that Petri nets can be viewed as a particular form of workflow).

[1] https://en.wikipedia.org/wiki/Excalibur_(film)

This line of research can clearly go much beyond in future more ambitious projects, in particular if the Petri net model is extended. In actual practice, as remarked later in his introductory process mining paper [Aalst], the event logs can register much richer information than the bare parameter-less traces:

> Moreover, event logs may contain information about resources, timestamps, and case data; for example, an event referring to activity "register request" and case "992564" may also have attributes describing the person registering the request (such as "John"), time of the event (such as "3011-2011:14.55"), age of the customer (such as "45"), and claimed amount (such as "650 euro"). After aligning model and log the event log can be replayed on the model; while replaying, one can analyze these additional attributes; [with the necessary additional information] it is possible to analyze wait times between activities. Measuring the time difference between causally related events and computing basic statistics (such as averages, variances, and confidence intervals) makes it possible to identify the main bottlenecks.

The present text is organized as follows. Sections 2 and 3 discuss, respectively, our approach to the situation calculus model and to the Petri net model, stressing their complementarity. In both sections the elementary request processing application is used as example. Section 4 exploits – again in terms of the two models – the structurally analogous digital entertaining application.

In full recognition of the distinctive characteristics of digital entertainment, wherein narration must visually show the participating characters enacting the events of the generated plots, which in turn should allow immersive user interaction, a fully implemented dramatization feature is described in detail in sub-section 4.3. Finally, concluding remarks are the object of section 5.

## 2. The Situation Calculus model

The starting point of our long-term **Logtell** project was the development of running conceptual level specifications to characterize narrative genres in a logic programming formalism [Furtado and Ciarlini]. To specify an information system application, it is not enough to define the classes of facts that will eventually populate its database component. One should specify, also in conceptual terms, i.e. in the language of the application domain, a fixed repertoire of events, whereby the state of the underlying mini-world would change.

Accordingly, besides our **static schema**, wherein facts are specified in terms of the Entity-Relationship model [Batini et al.], we provided a **dynamic schema** to describe a fixed repertoire of event-producing operations for performing state changes in conformance with the applicable integrity constraints. Following the **S**tanford **R**esearch **I**nstitute **P**roblem **S**olver (**STRIPS**) method [Fikes and Nilsson], each operation is defined in terms of *pre-conditions*, which consist of conjunctions of positive and/or negative terms expressing facts, and any number of post-conditions, consisting of facts to be *asserted* or *retracted* as the effect of executing the operation.

This approach is fully consistent with the *situation calculus* theory, compactly expressed as follows [Kowalski]: “The following two clauses constitute the core of our formulation of the situation calculus as a logic program. Together they define what sentences **P** hold in the situation **result(A, S)** that is the result of the transition from **state S** by an **action of type A**:

holds(P, result(A, S)) ← happens(A, S) ∧ initiates(A, S, P)
holds(P, result(A, S)) ← happens(A, S) ∧ holds(P, S) ∧ ¬ terminates(A, S, P)”

noting that the second clause of this second-order logic formulation avoids the exponential proliferation of first-order logic clauses constituting the so-called *frame problem*, by eliminating the need to specify every class of facts (P) that are *not* affected by the execution of an operation (A).

In turn, these situation calculus clauses suggest an elementary plan-generator, which can be thus expressed in natural language:

- A fact F holds if it is true in the initial state
- F holds if it is **added** as one of the effects of an operation Op, and the **preconditions** of Op hold at the current state
- F holds after the execution of an operation Op if it did already hold at the current state and if it is not **deleted** as one of the effects of Op

and then translated into a Prolog program:

```
holds(Fact, [start]) :-
  initial_state(Fact), !.

holds(Fact,[Operation|Current_state]) :-
  added(Fact, Operation),
  precond(Operation, Current_state).

holds(Fact,[Operation|Current_state]) :-
  not deleted(Fact, Operation),
  holds(Fact, Current_state).
```

Although this elementary program is able to handle overly simple cases, such as the well-known monkey-and-bananas problem, it must be considerably expanded for practical usage. With that in mind, we developed two different Prolog programs for plan generation, both obeying the *backward-chaining* regime of the elementary program, one in Sicstus Prolog and the other in SWI Prolog [Ciarlini et al. 2005; Ciarlini et al. 2009]. We also provided a simulated execution feature, which follows a given plan in *forward-chaining* direction, testing the preconditions of the successive operations and performing additions and deletions to achieve their postconditions.

Planners can be used by *agents*, in *situations* involving either opportunities or risks to them, in order to reach some *goal* in a new target state where such opportunities are materialized and/or any risks avoided. In business applications the generated plans correspond to potential use cases, whereas in digital entertainment they represent automatically composed *narrative plots*.

The pragmatic aspect introduced by relating agents, situations and goals leads to the specification of a third schema (not to be treated in this paper – cf. [Ciarlini et al. 2005]), the *behavioural schema*, which in its minimal version, consists of *goal-inference rules*.

As illustration, we shall concentrate on a *request processing* example, used to introduce the fundaments of the process mining project initiated by Wil van der Aalst [Aalst], as informally described in the paragraph below, showing in blue the kinds of events that can take place:

> The **client** of a business company **registers** a **request**, asking for the compensation of a certain value. To serve as a document to justify the claim, he encloses a **ticket**. The request itself is either **examined thoroughly** or **examined casually**, and the ticket is separately **checked**. After that, the company **decides** either to **pay the compensation** or to **reject the request**. It is also possible, provided that neither the payment nor the rejection have been executed, to **reinitiate the process** utilizing the same registered information.

The entities involved and their properties are specified, as explained before, through a **static schema**:

```
entity(request, rn).
attribute(request, r_value).
attribute(request, examined).
attribute(request, analyzed).

entity(client, cn).
attribute(client, status).

entity(ticket, tn).
attribute(ticket, t_value).
attribute(ticket, checked).

relationship(claims,
    [client, request]).
attribute(claims, payed).
attribute(claims, rejected).

relationship(is_holder,
    [client, ticket]).

relationship(attached_to,
     [ticket, request]).
```

and the event-producing operations in a **dynamic schema:**

```
operation(register(C,V,T,R)).
precond(register(C,V,T,R), (client(C), ticket(T),
atom_concat(req_,T,R))).
added(request(R), register(C,V,T,R)).
added(claims(C,R), register(C,V,T,R)).
added(r_value(R,V), register(C,V,T,R)).
added(attached_to(T,R), register(C,V,T,R)).

operation(examine_thoroughly(R,C)).
precond(examine_thoroughly(R,C), (claims(C,R), request(R), client(C),
status(C,ok))).
added(examined(R,C), examine_thoroughly(R,C)).

operation(examine_casually(R,C)).
precond(examine_casually(R,C), (claims(C,R), request(R), client(C))).
added(examined(R,C), examine_casually(R,C)).

operation(check_ticket(R,C,T)).
precond(check_ticket(R,C,T), (claims(C,R),is_holder(C,T),
attached_to(T,R), r_value(R,V),  t_value(T,V))).
added(checked(T,R), check_ticket(R,C,T)).

operation(decide(R,C,V,D)).
precond(decide(R,C,V,D), (examined(R,C), attached_to(T,R), checked(T,R),
r_value(R,V),
   if(V =< L, D = ok, D = (not ok)))) :- limit(L).
added(analyzed(R,D), decide(R,C,V,D)).
```

```
operation(pay_compensation(R,C,V)).
precond(pay_compensation(R,C,V), (claims(C,R), analyzed(R, D),
r_value(R,V), D = ok)).
added(payed([C,R],V), pay_compensation(R,C,V)).

operation(reject_request(R,C,V)).
precond(reject_request(R,C,V), (claims(C,R), analyzed(R, D),
   r_value(R,V), D = (not ok))).
added(rejected([C,R],'limit exceeded'), reject_request(R,C,V)).

operation(reinitiate_request(R,C,T,V)).
precond(reinitiate_request(R,C,T,V), (analyzed(R,D),claims(C,
R),t_value(T,V),
    not payed([C,R],V),not rejected([C,R],M))).
deleted(examined(R,C),reinitiate_request(R,C,T,V)).
deleted(checked(T,R),reinitiate_request(R,C,T,V)).
deleted(analyzed(R,D),reinitiate_request(R,C,T,V)).
```

In order to conduct experiments, an **initial state** must be introduced, indicating the instances of the entity classes and the specific initial values of their properties, as well as a clause (`limit`) affecting the precondition of the **decide** operation:

```
limit(100).

client('Mary').
status('Mary',ok).
ticket(t123).
t_value(t123, 58).
is_holder('Mary',t123).

client('Peter').
status('Peter',overdue).
ticket(t124).
t_value(t124, 200).
is_holder('Peter',t124).
```

To test the specification via the plan-generator, on the basis of the declared initial state, a `try_request` predicate was provided. Its first clause tries the favorable outcome that constitutes the client's goal. If this fails, the negative outcome is treated by the second clause, which exposes the motive of rejection:

```
try_request(C,V,T,R, P) :-
  plans((claims(C,R), r_value(R,V),payed([C,R],V)), P),
  nl, write('Value credited: '), write(V), nl.
try_request(C,V,T,R, P) :-
  plans((claims(C,R),r_value(R,V),rejected([C,R],M)),P),
  nl,  write('Motive of rejection: '),  write(M), nl.
```

As a consequence of the values attributed in the initial state, client Mary succeeds whereas client Peter, who is applying for a reimbursement that exceeds the prescribed limit, has his request denied. By repeatedly entering ";", the user is allowed to view alternative plans, choosing either `examine_thoroughly` or `examine_casually`, and letting `check_ticket` either precede or follow the chosen examination action. Note that Peter's plans, because of his 'overdue' status, do not include the `examine_thoroughly` option.

```
?- try_request('Mary', 58, t123, R, P).
```

```
Value credited: 58
R = req_t123,
P = start=> register('Mary', 58, t123, req_t123) =>
examine_thoroughly(req_t123, 'Mary') => check_ticket(req_t123, 'Mary',
t123) => decide(req_t123, 'Mary', 58, ok) => pay_compensation(req_t123,
'Mary', 58)  ;

Value credited: 58
R = req_t123,
P = start => register('Mary', 58, t123, req_t123) =>
check_ticket(req_t123, 'Mary', t123) => examine_thoroughly(req_t123,
'Mary') => decide(req_t123, 'Mary', 58, ok) =>
pay_compensation(req_t123, 'Mary', 58)  ;

Value credited: 58
R = req_t123,
P = start=>register('Mary', 58, t123,
req_t123)=>examine_casually(req_t123, 'Mary')=>check_ticket(req_t123,
'Mary', t123)=>decide(req_t123, 'Mary', 58,
ok)=>pay_compensation(req_t123, 'Mary', 58)  ;

Value credited: 58
R = req_t123,
P = start=>register('Mary', 58, t123, req_t123) => check_ticket(req_t123,
'Mary', t123)=>examine_casually(req_t123, 'Mary')=> decide(req_t123,
'Mary', 58, ok)=>pay_compensation(req_t123, 'Mary', 58)  ;

?- try_request('Peter', 200, t124, R, P).
Motive of rejection: limit exceeded
R = req_t124,
P = start => register('Peter', 200, t124, req_t124) =>
examine_casually(req_t124, 'Peter') => check_ticket(req_t124, 'Peter',
t124) => decide(req_t124, 'Peter', 200, not ok) =>
reject_request(req_t124, 'Peter', 200)  ;

Motive of rejection: limit exceeded
R = req_t124,
P = start => register('Peter', 200, t124, req_t124) =>
check_ticket(req_t124, 'Peter', t124) => examine_casually(req_t124,
'Peter') => decide(req_t124, 'Peter', 200, not ok) =>
reject_request(req_t124, 'Peter', 200)  ;
```

In turn, one of the tasks made possible by the simulated execution feature is checking plan validity. We utilized it in a `check_fix` predicate to verify whether a manually created plan is valid. The simulated execution proceeds as long as the given plan is correct, and pauses at the point where an error is detected. For the current version of `check_fix`, corrections are provided in two simple error cases: missing operations to fulfill preconditions and the presence of redundant operations. Both problems occur in the example below. Note that, once a correction is introduced, the result is treated again as a given plan, the process being repeated until the predicate returns a **Valid** assessment.

```
read(P), check_fix(P,Pv),!.

given plan:
start=>register(Peter,200,t124,req_t124)=>decide(req_t124,Peter,200,_506)
=>examine_casually(req_t124,Peter)=>reject_request(req_t124,Peter,200)
```

```
not enabled: decide(req_t124,Peter,200,_506)

plan with correction:
start=>register(Peter,200,t124,req_t124)=>examine_casually(req_t124,Peter
)=>check_ticket(req_t124,Peter,t124)=>decide(req_t124,Peter,200,not
ok)=>examine_casually(req_t124,Peter)=>reject_request(req_t124,Peter,200)

given plan:
start=>register(Peter,200,t124,req_t124)=>examine_casually(req_t124,Peter
)=>check_ticket(req_t124,Peter,t124)=>decide(req_t124,Peter,200,not
ok)=>examine_casually(req_t124,Peter)=>reject_request(req_t124,Peter,200)

redundant: examine_casually(req_t124,Peter)

plan with correction:
start=>register(Peter,200,t124,req_t124)=>examine_casually(req_t124,Peter
)=>check_ticket(req_t124,Peter,t124)=>decide(req_t124,Peter,200,not
ok)=>reject_request(req_t124,Peter,200)

Valid
```

Notice that, when inserting the operations needed to fulfill preconditions of the `decide` operation, the `examine_casually` option was chosen by the simulation algorithm, because the other option, `examine_thoroughly` – which would be considered first in consequence of its preceding position in the dynamic schema – was, as explained before, not available to Peter due to his status.

## 3. The Petri net model

In database practice, *Process Mining* [Aalst] has been increasingly gaining attention as a vitally important discipline. It relies on the analysis of an adequately structured *execution log* to check the compliance of the users' transactions, as represented by *traces* extracted from the execution log, with the intended behaviour of the application. In the unfortunately frequent case of *legacy systems*, in which no duly documented specification is available, Process Mining offers methods to discover what might be called an *implicit specification*, compatible with the observed traces.

As a formalism especially attractive for offering visual intuition, *Petri nets*, is commonly utilized in such methods. In the semi-formal terminology employed in this paper, a Petri net is a graph with two kinds of nodes: **places** (round nodes, either empty or containing exactly one **token**) and **transitions** (square nodes representing operations).

We define a Petri net **edge** as triple, **Op1** **Pn** **OP2**, with two (operational) transition nodes and an intervening place node**.** Petri net edges are positioned so as to express a partial ordering in the execution of events, which may follow each other in linear sequences, possibly branching to form and-forks, or-forks, and-joins, and or-joins. We shall also consider a simple cases of backward loop, allowing to return to a previous position and try different branching options.

Process mining promotes the generation of Petri nets by analyzing the order of events in the traces recorded in the execution log. Here we shall try a different approach, showing how the ordering requirements can be derived from the situation calculus model.

The first basic consideration is that there exists an edge connecting (through a **Pn** node) **Op1** and **Op2** if the post-conditions of **Op1** have a non-empty intersection with the pre-conditions of **Op2.** There is also an edge from **Op1** to **Op2** if some post-condition of **Op1** cancels some post-condition of **Op2** (thus causing a backward loop, whereby **Op2** can be retried).

*Forks* occur when **there** are edges leading from an **Op** node into two or more nodes **Op1**, **Op2**, …, **Opn**. A fork is an *or-fork* if **Op1**, **Op2**, …, **Opn** contain either *incompatible* pre-conditions or *redundant* post-conditions. Incompatibility typically results from conflicting value comparisons as well as from logic opposition (P *vs*. not P). In addition, we consider any pair **Opi** and **Opj** incompatible if the execution of one of them would be rendered impossible by the execution of the other, which might happen if the pre-conditions of one of these operations require P (or not P), whereas the post-conditions of the other produce the deletion of P (or, respectively, the addition of P). A fork is an *and-fork* if none of these situations holds.

*Joins* occur when there are edges from two or more nodes **Op1**, **Op2**, …, **Opn** into a single node **Op**. A join is an or-join if **Op1**, **Op2**, …, **Opn** contain *redundant* post-conditions, or post-conditions that cancel the post-conditions of **Op** (which induces a backward loop enabling **Op** to be retried). A join is an *and-join* if none of these situations holds.

By thus considering the presence of edges, as well forks and joins, we have a method to generate a clausal representation of the Petri net corresponding to a given situation calculus specification.

To illustrate the result of the method, we shall return to the previously specified request processing application which, as remarked in the Introduction, Wil van des Aalst employed as an elementary introductory example [Aalst].

In that work, as a convenient abbreviation, the operations are represented by a single character: **a** = register request, **b** = examine thoroughly, **c** = examine casually, **d** = check ticket, **e** = decide, **f** = reinitiate request, **g** = pay compensation, and **h** = reject request. With this compact notation, the trace **acdeh** describes a reimbursement request that was ultimately rejected: **a**:registration, **c**:casual examination, **d**:check, **e**:decision, **h**:reject. Figure 1 displays the execution log and the Petri net itself.

event log

| # | trace |
|---|---|
| 455 | acdeh |
| 191 | abdeg |
| 177 | adceh |
| 144 | abdeh |
| 111 | acdeg |
| 82 | adceg |
| 56 | adbeh |
| 47 | acdefdbeh |
| 38 | adbeg |
| 33 | acdefbdeh |
| 14 | acdefbdeg |
| 11 | acdefdbeg |
| 9 | adcefcdeh |
| 8 | adcefdbeh |
| 5 | adcefbdeg |
| 3 | acdefbdefdbeg |
| 2 | adcefdbeg |
| 2 | adcefbdefbdeg |
| 1 | adcefdbefbdeh |
| 1 | adbefbdefdbeg |
| 1 | adcefdbefcdefdbeg |
| 1391 | |

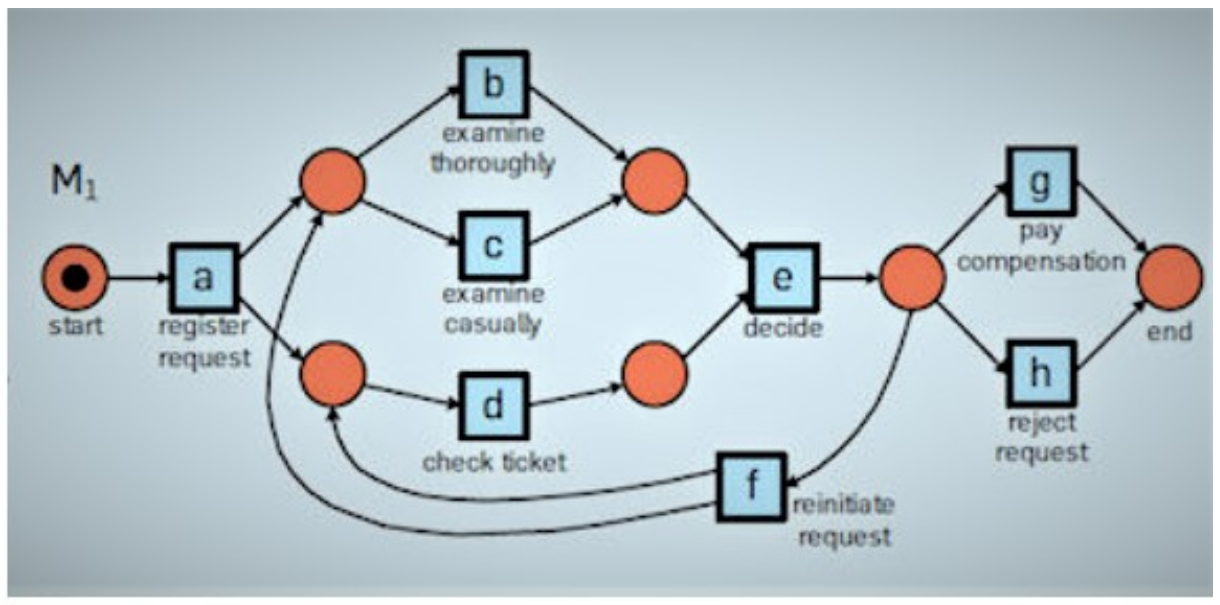

Figure 1: The Wil van des Aalst's example

By applying our method, we obtain the clausal representation below, where the one letter abbreviations appear as operation labels:

```
start - a:register(c,v,t,r)
a:register(c,v,t,r)             - s(1) - b:examine_thoroughly(r,c)
a:register(c,v,t,r)             - s(1) - c:examine_casually(r,c)
a:register(c,v,t,r)             - s(2) - d:check_ticket(r,c,t)
b:examine_thoroughly(r,c)       - s(3) - e:decide(r,c,v,d)
c:examine_casually(r,c)         - s(3) - e:decide(r,c,v,d)
d:check_ticket(r,c,t)           - s(4) - e:decide(r,c,v,d)
e:decide(r,c,v,d)               - s(5) - f:reinitiate_request(r,c,t,v)
e:decide(r,c,v,d)               - s(5) - g:pay_compensation(r,c,v)
e:decide(r,c,v,d)               - s(5) - h:reject_request(r,c,v)
f:reinitiate_request(r,c,t,v) - s(1) - b:examine_thoroughly(r,c)
f:reinitiate_request(r,c,t,v) - s(1) - c:examine_casually(r,c)
f:reinitiate_request(r,c,t,v) - s(2) - d:check_ticket(r,c,t)
g:pay_compensation(r,c,v) - end
h:reject_request(r,c,v)   - end
```

The detected cases of or-forks and join-forks are listed below, using bold to distinguish the **Op** originating / terminating operation:

**Or-forks**

**register(c,v,t,r)** - examine_casually(r,c)- examine_thoroughly(r,c)
**decide(r,c,v,d)** - pay_compensation(r,c,v)- reinitiate_request(r,c,t,v)
**decide(r,c,v,d)** - pay_compensation(r,c,v)- reject_request(r,c,v), reinitiate_request(r,c,t,v)
**reinitiate_request**(r,c,t,v) - examine_casually(r,c)- examine_thoroughly(r,c)

**Join-forks**

register(c,v,t,r) - reinitiate_request(r,c,t,v)- **examine_thoroughly(r,c)**
register(c,v,t,r) - reinitiate_request(r,c,t,v)- **examine_casually(r,c)**
register(c,v,t,r) - reinitiate_request(r,c,t,v)- **check_ticket(r,c,t)**

Operations coming from and-forks can be executed in any order. They might be executed in parallel if there are enough resources, but we shall confine ourselves to linear sequences.

Once the clausal representation is generated, it is possible to display the Petri net (via the DrawNetwork utility), as shown in Figure 2. Placing the cursor over a node with a label such as **e**, causes the operation corresponding to this label, namely **decide**, to be shown.

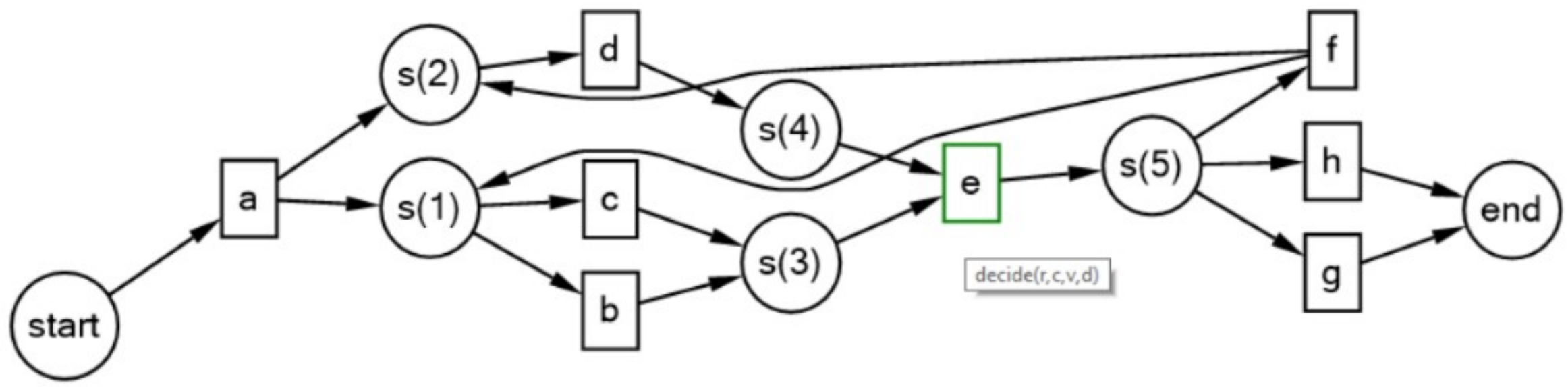

Figure 2: Petri net drawn from the request processing clausal representation

Even more importantly, one can check whether a given trace fits the specified ordering, as a linear sequence initiating in the start node and terminating in the end node.

As a preliminary consideration, note that, by construction, or-forks stem from place nodes, and or-joins always converge to a single place node. Since place nodes can contain at most one token, or-type branching is restricted – as should be expected – to the selection of a single option.

In contrast, and-forks stem from operation nodes, and and-joins converge to an operation node. Differently from place nodes, operation nodes are able to emit tokens to all outgoing place nodes (one token for each).

According to the standard token-based Petri net process, forming a trace begins by placing a token in the start place node. The place node is then activated which signifies that the token is consumed and the single operation node attached to the start node is enabled. In the next steps, successive place nodes are activated after receiving tokens from enabled nodes, and some operation node to which all incoming place nodes are active (i.e. contain a token) is chosen to be enabled. The process ends successfully when some operation node connected to the end place node is reached.

The Prolog program below allows to check whether a given trace, such as **`acdefdbeg`** is valid, using a compact version of the clausal representation. Notice in the recursive `check1` subprogram the alternation between `activate` and `enable` tasks.

```
check(T) :-
  gen_arcs,
  check1(T),
  enabled(E),
  e_arc(E,end).

check1(T) :-
  i_arc(start,I),
  assert(enabled(I)),
  atom_chars(T,[I|ES]),
  ch1(ES).

ch1([]) :- !.
ch1([E|R]) :-
  activate(S),
  enable(S,E),
  ch1(R).

activate(S) :-
  setof(Se,
        Ei^(enabled(E),arc(E,Se,Ei)),
        Ss),
```

```
i_arc(start, a).
arc(a, s(1), b).
arc(a, s(1), c).
arc(a, s(2), d).
arc(b, s(3), e).
arc(c, s(3), e).
arc(d, s(4), e).
arc(e, s(5), f).
arc(e, s(5), g).
arc(e, s(5), h).
arc(f, s(1), b).
arc(f, s(1), c).
arc(f, s(2), d).
e_arc(g, end).
e_arc(h, end).
```

```
  forall(member(Se,Ss),
         assert(activated(Se))),
  setof(Se,activated(Se),S).

enable(S,Ei) :-
  retract(enabled(A)),
  forall(arc(Ej,Si,Ei),
         member(Si,S)),
  assert(enabled(Ei)),
  forall((member(Si,S),once(arc(Ej,Si,Ei))),
         retract(activated(Si))).
```

**`?- check(acdefdbeg).`**
**`True`**

Also available (and completely reproduced in the Appendix, so that the reader can play with it if SWI-Prolog is installed) is an interactive trace-generation program, `traverse` which, at each step, allows the user to choose from the current options. Whenever only one option is applicable it is automatically inserted. An example run follows:

```
?- traverse.
a
choose one label from:
b:examine_thoroughly
c:examine_casually
d:check_ticket
my choice: c
d
e
choose one label from:
f:reinitiate_request
g:pay_compensation
h:reject_request
my choice: f
choose one label from:
b:examine_thoroughly
c:examine_casually
d:check_ticket
my choice: d

choose one label from:
b:examine_thoroughly
c:examine_casually
my choice: b
e
choose one label from:
f:reinitiate_request
g:pay_compensation
h:reject_request
my choice: g

acdefdbeg

start=>register(c,v,t,r)=>examine_casually(r,c)=>check_ticket(r,c,t)=>dec
ide(r,c,v,d)=>reinitiate_request(r,c,t,v)=>check_ticket(r,c,t)=>examine_t
horoughly(r,c)=>decide(r,c,v,d)=>pay_compensation(r,c,v)
```

## 4. A digital entertainment application

Our proposed complementary approach is equally suitable to narratives composed for entertainment. Here we shall adapt an incident taken from the film *Excalibur*, directed, produced, and co-written by John Boorman in 1981. The incident can be thus summarized:

> Sir Gawain accuses Queen Guinevere of adultery (figure 3). A trial by combat is announced, there being two candidate knights to claim the Queen's innocence: Sir Lancelot, a worthy knight, famous for his many victories, and Sir Perceval, who would be no less reputed in the future, but at that time still had little combat experience. The Queen would be vindicated if her defender could defeat the accuser, otherwise she would be condemned. The trial could be reinitiated if a last minute replacement of defender chanced to occur.

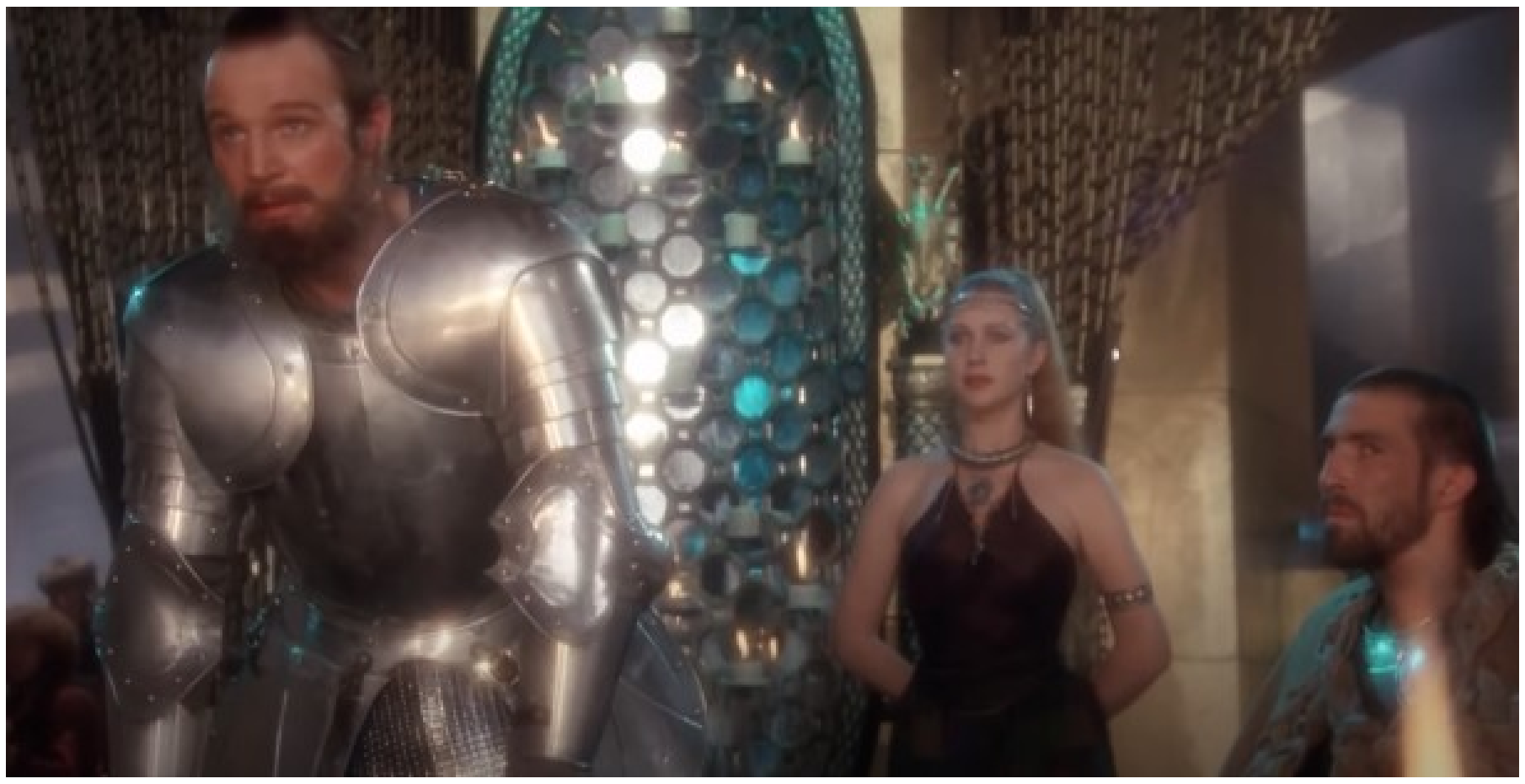

Figure 3: Sir Gawain denounces the Queen

### 4.1. Trial by combat in the situation calculus model

We start by providing a specification modelled on our situation calculus formalism, with static and dynamic schemas, followed by the initial state that was used to exploit the specification.

### Static schema

```
entity(person, pn).

entity(knight, kn).
attribute(knight, strength).
attribute(knight, loyal).

entity(accuser, an).

entity(defendant, dn).
attribute(defendant,has_defender).

entity(defender, dn).

entity(challenger, cn).

entity(offense,on).
```

```
relationship(accusation,
     [defendant, offense]).
attribute(accusation, vindicated).
attribute(accusation, condemned).

relationship(encounter,
     [challenger, defender]).
attribute(encounter, winner).
```

## Dynamic schema

```
operation(accuse(A,D,O)).
precond(accuse(A,D,O), (knight(A), loyal(A,false), person(D),
offense(O))).
added(accusation(D,O), accuse(A,D,O)).
added(accuser(A), accuse(A,D,O)).
added(defendant(D), accuse(A,D,O)).

operation(enter_worthy_defender(K,D,O)).
precond(enter_worthy_defender(K,D,O), (accusation(D,O), knight(K),
loyal(K,true), strength(K,S), S > 100)).
added(has_defender(D,true), enter_worthy_defender(K,D,O)).
added(defender(K), enter_worthy_defender(K,D,O)).

operation(enter_beginner_defender(K,D,O)).
precond(enter_beginner_defender(K,D,O), (accusation(D,O), knight(K),
loyal(K,true), strength(K,S), S =< 100)).
added(has_defender(D,true), enter_beginner_defender(K,D,O)).
added(defender(K), enter_beginner_defender(K,D,O)).

operation(enter_challenger(A,D,O)).
precond(enter_challenger(A,D,O), (accuser(A), accusation(D,O))).
added(challenger(A), enter_challenger(A,D,O)).

operation(combat(A,K,D,O,V)).
precond(combat(A,K,D,O,V), (accusation(D,O), challenger(A),
defender(K),strength(K,Sk), strength(A,Sa), if(Sk > Sa,V = K, V = A))).
added(encounter(A,K), combat(A,K,D,O,V)).
added(winner([A,K],V), combat(A,K,D,O,V)).

operation(vindicate(D,O)).
precond(vindicate(D,O), (accusation(D,O), challenger(A), defender(K),
winner([A,K],V), V = K)).
added(vindicated([D,O],innocent), vindicate(D,O)).

operation(condemn(D,O)).
precond(condemn(D,O), (accusation(D,O), challenger(A), defender(K),
winner([A,K],V), not (V = K))).
added(condemned([D,O],guilty), condemn(D,O)).

operation(reinitiate_trial(A,K,D,O,V)).
precond(reinitiate_trial(A,K,D,O,V),(accusation(D,O),encounter(A,K),winne
r([A,K],V),not vindicated([D,O],innocent), not condemned([D,O],guilty))).
deleted(encounter(A,K),reinitiate_trial(A,K,D,O,V)).
deleted(winner([A,K],V),reinitiate_trial(A,K,D,O,V)).
deleted(challenger(A),reinitiate_trial(A,K,D,O,V)).
deleted(defender(K),reinitiate_trial(A,K,D,O,V)).
deleted(has_defender(D,true),reinitiate_trial(A,K,D,O,V)).
```

**Initial state**

```
person('Guinevere').

knight('Lancelot').
loyal('Lancelot',true).
strength('Lancelot',200).

knight('Perceval').
loyal('Perceval',true).
strength('Perceval',100).

knight('Gawain').
loyal('Gawain',false).
strength('Gawain',150).

offense(murder).

offense(adultery).
```

Described in this fashion, the well-intentioned but still immature Perceval would stand no chance to defeat Gawain, momentarily playing the role of a villain.

```
?- test_Perceval.
start=>accuse(Gawain,Guinevere,adultery)=>enter_challenger(Gawain,Guineve
re, adultery)=>enter_beginner_defender(Perceval,Guinevere,adultery)=>
combat(Gawain,Perceval,Guinevere,adultery,Gawain)=>
condemn(Guinevere,adultery)
```

On the contrary, Lancelot had what was required to triumph, thereby establishing the Queen's innocence (figure 4).

```
?- test_Lancelot.
start=>accuse(Gawain,Guinevere,adultery)=>enter_challenger(Gawain,Guineve
re, adultery)=>enter_worthy_defender(Lancelot,Guinevere,adultery)=>
combat(Gawain,Lancelot,Guinevere,adultery,Lancelot)=>
vindicate(Guinevere,adultery)
```

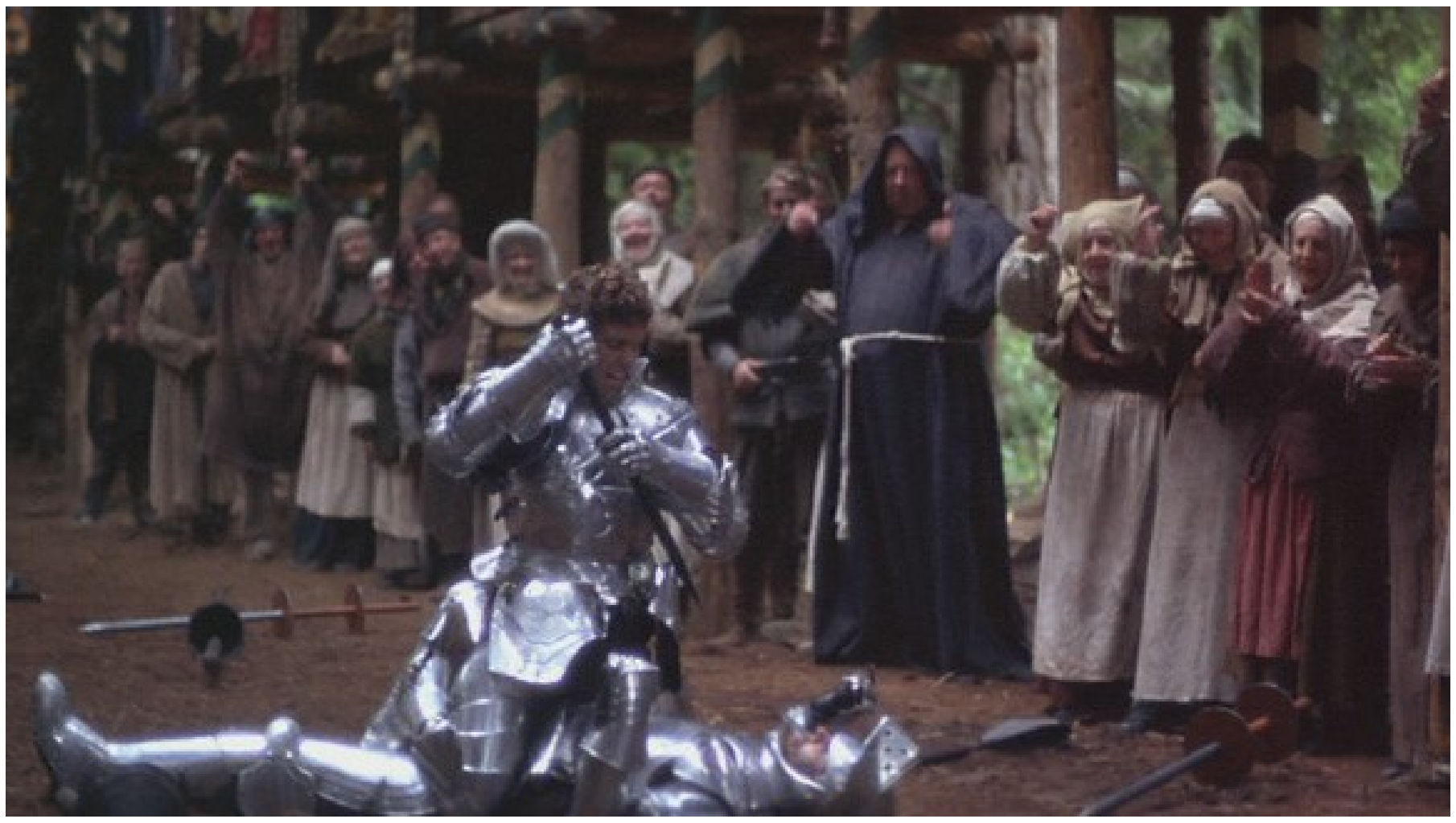

Figure 4: the Queen is innocent!

## 4.2. Trial by combat in the Petri net model

It so happens that, when mapped into a Petri net representation, the trial by combat digital entertainment example exhibits the same structure of the business-oriented request processing application. Therefore, for the sake of brevity, we shall simply list, trusting that the reader will have no difficulty to interpret, the clausal representation, the generated Petri net image (figure 5), and the result of executing the traverse predicate – at each user interaction providing the same choices as when applying the predicate to the 'serious' request processing application. As expected, the two traces are identical, though corresponding to plans staged in quite different domains.

### Clausal representation

```
start - a:accuse(a,d,o)
a:accuse(a,d,o)                    - s(1) - b:enter_worthy_defender(k,d,o)
a:accuse(a,d,o)                    - s(1) - c:enter_beginner_defender(k,d,o)
a:accuse(a,d,o)                    - s(2) - d:enter_challenger(a,d,o)
b:enter_worthy_defender(k,d,o)     - s(3) - e:combat(a,k,d,o,v)
c:enter_beginner_defender(k,d,o)   - s(3) - e:combat(a,k,d,o,v)
d:enter_challenger(a,d,o)          - s(4) - e:combat(a,k,d,o,v)
e:combat(a,k,d,o,v)                - s(5) - f:reinitiate_trial(a,k,d,o,v)
e:combat(a,k,d,o,v)                - s(5) - g:vindicate(d,o)
e:combat(a,k,d,o,v)                - s(5) - h:condemn(d,o)
f:reinitiate_trial(a,k,d,o,v)      - s(1) - b:enter_worthy_defender(k,d,o)
f:reinitiate_trial(a,k,d,o,v)      - s(1) - c:enter_beginner_defender(k,d,o)
f:reinitiate_trial(a,k,d,o,v)      - s(2) - d:enter_challenger(a,d,o)
g:vindicate(d,o) - end
h:condemn(d,o) - end
```

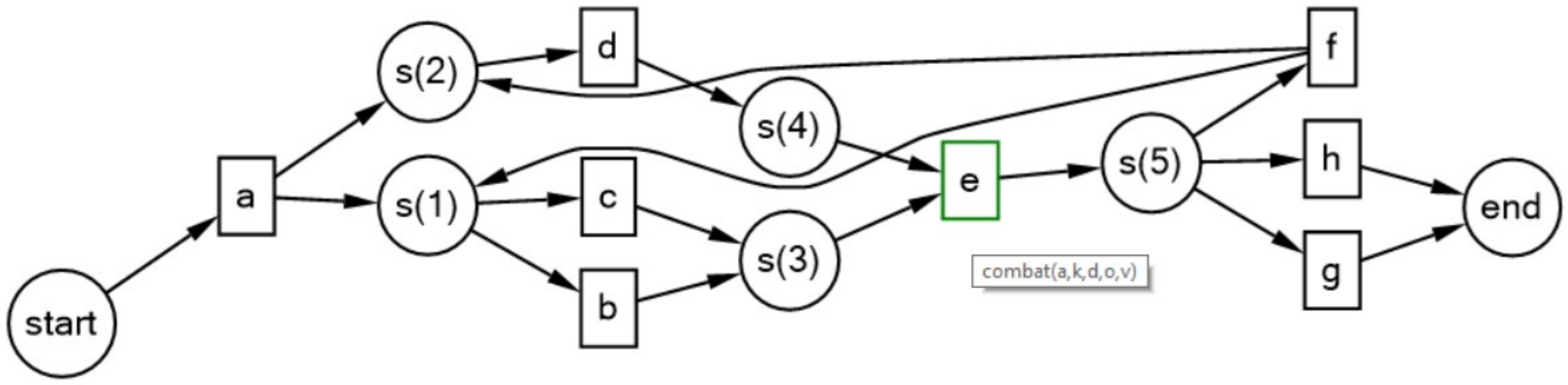

Figure 5: Petri net drawn from the trial by combat clausal representation

### Interactive trace generation

```
:- traverse.

a
choose one label from:
b:enter_worthy_defender
```

```
c:enter_beginner_defender
d:enter_challenger

my choice: c
d
e
choose one label from:
f:reinitiate_trial
g:vindicate
h:condemn
my choice: f
choose one label from:
b:enter_worthy_defender
c:enter_beginner_defender
d:enter_challenger
my choice: d

choose one label from:
b:enter_worthy_defender
c:enter_beginner_defender
my choice: b
e
choose one label from:
f:reinitiate_trial
g:vindicate
h:condemn
my choice: g

acdefdbeg

start=>accuse(a,d,o)=>enter_beginner_defender(k,d,o)=>enter_challenger(a,
d,o)=>combat(a,k,d,o,v)=>reinitiate_trial(a,k,d,o,v)=>enter_challenger(a,
d,o)=>enter_worthy_defender(k,d,o)=>combat(a,k,d,o,v)=>vindicate(d,o)
```

### 4.3. Dramatization of the Trial by combat using the Petri net model

Generating an entire plot via the traverse call may give an indication of the effectiveness of our method to generate interactive stories, but is not wholly satisfactory from a digital entertainment viewpoint. In general, interactive storytelling systems are composed of two main modules: a plot-generator and a dramatization module. While the plot-generator handles the logical process of creating interactive narratives, the dramatization module translates the events of the plot into a form of visual representation. In the course of our Logtell project,[2] we developed several dramatization modules to allow users to visualize narratives in different formats, including frame picture sequences in storyboard style [Ciarlini et al. 2009], comics book formats [Barbosa et al.; Lima et al. 2013], 3D graphics [Ciarlini et al. 2005; Lima 2010], augmented reality [Lima et al. 2014a; Franco and Lima], video-based techniques [Lima et al. 2012; Lima 2014b; Lima et al. 2018a], and game-based formats [Lima et al. 2014; Lima et al. 2018b]. In this work, we present a dramatization module capable of representing the generated Petri net model using 2D graphics and animations.

The architecture of our interactive storytelling system is based on a client-server model (Figure 6), where the server is responsible for the generation of the plot (Petri net model) and the client handles the dramatization. On the server-side, the Network Manager receives plot requests from clients and uses the Prolog implementation described in the previous sections to generate a Petri net, which is then sent to clients for dramatization. On the client-side, the Drama Manager interprets and controls the execution of the Petri net by sending action requests to virtual Actors. The process of composing scenes for dramatization (i.e., selecting the Actors and Locations to show) is performed by the Scene Composer, which is constantly being informed by the Drama Manager about the type of scene being dramatized.

[2] http://www.icad.puc-rio.br/~logtell/

User interaction is handled by the Interaction App module, which is implemented as a mobile app that uses a Convolutional Neural Network classifier to identify hand-draw sketches (see [Lima et al. 2020] for more details about the sketch recognition process). Once a sketch is recognized, its identification class is sent to the Interaction Server through a TCP/IP network message. The Interaction Server module is responsible for receiving and interpreting the sketch classes sent by clients. Two interaction modes are supported: (1) single user mode, in which the first valid user sketch received by the system is immediately used as the interference choice be incorporated into the story; and (2) voting mode, in which the Interaction Server collects all users' sketches during a certain time and then selects one through a voting process.

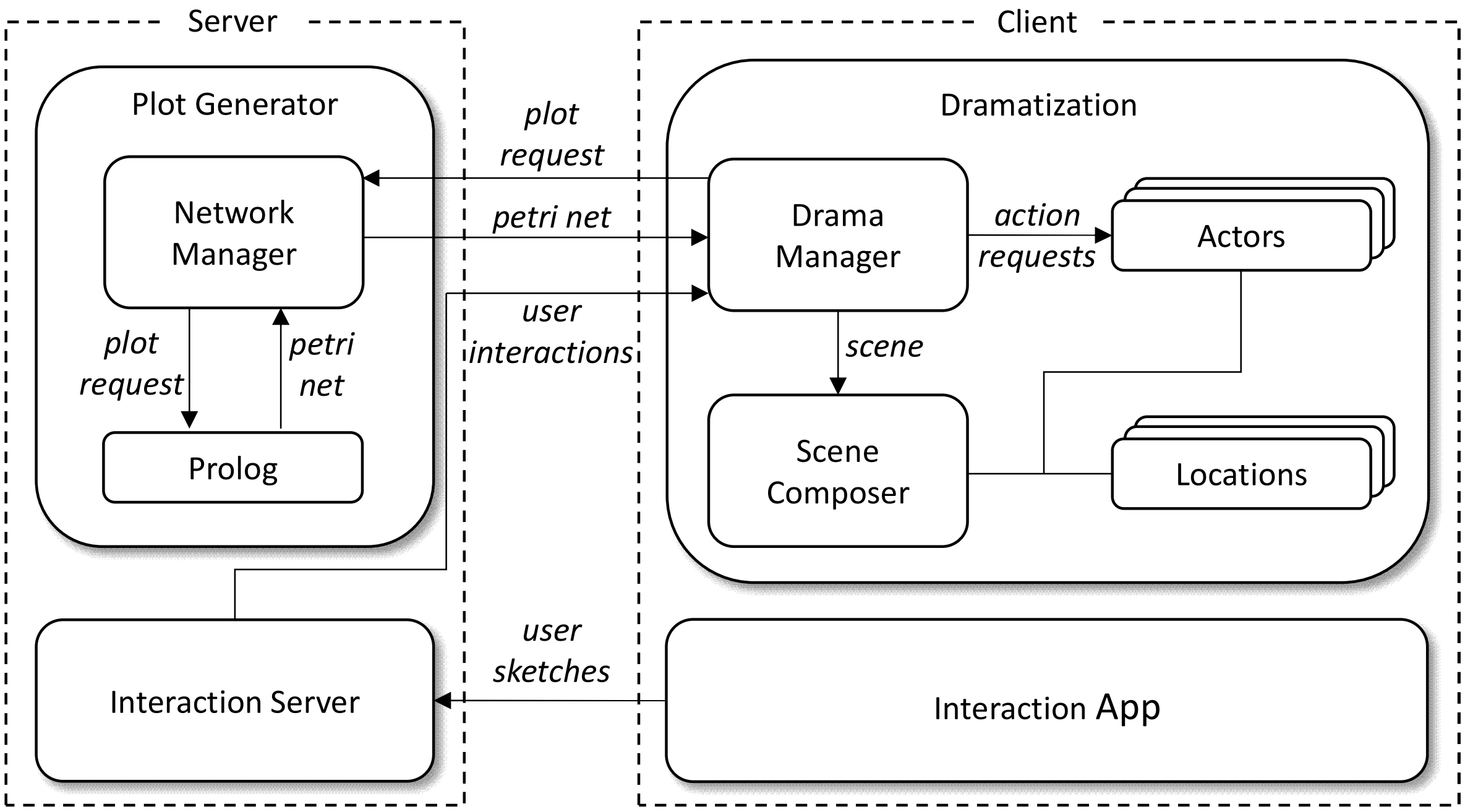

Figure 6: Architecture of our interactive storytelling system

Multiple programming languages were used in the implementation of our interactive storytelling system. As described in the previous sections, the process of generating plots in the Petri net model is implemented in Prolog. However, the Plot Generator also includes an additional module called Network Manager, which is implemented in C# and provides network communication capabilities to the system, allowing us to implement the plot generation process as a service provided by a network server. On the client-side, the dramatization system is implemented in Lua[3] using the Löve 2D framework,[4] which provides the graphical functionalities needed to create visual representations for the story. The interaction process is implemented in Java as an Android app, which communicates with a web service implemented in PHP. See [Gheno et al.] for more details about the design of the sketch-based interaction system.

[3] Lua is a well-known programming language developed at the Pontifical Catholic University of Rio de Janeiro, Brazil (http://www.lua.org/).

The Petri net representation of the plot created by the Plot Generator consists of a directed graph $G = (V, E)$, where $V$ is a set of nodes $\{v_1, v_2, \ldots, v_n\}$ and $E$ is a set of edges $\{e_i = (v_i, v_j), \ldots, e_m = (v_k, v_w)\}$. Each node $v_i$ is a pair $(id_i, ev_i)$, where $id_i$ is a unique name that identifies the node $v_i$ and $ev_i$ is an event description in a predicate format for *transition* nodes (e.g., `accuse(a,d,o)`), or the constant `nil` for *place* nodes (as described in Section 3, places are nodes that can contain tokens and transitions are nodes that represent operations).

When encoding the Petri net to be sent to the dramatization system, the graph is simplified as a set of edges, where each edge is represented in the format $[id_i : ev_i , id_j : ev_j]$. For example, the Petri net for the trial by combat (Figure 5) can be described as follows:

```
[start:nil, a:accuse(a,d,o)]
[a:accuse(a,d,o), s(1):nil]
[a:accuse(a,d,o), s(2):nil]
[b:enter_worthy_defender(k,d,o), s(3):nil]
[c:enter_beginner_defender(k,d,o), s(3):nil]
[d:enter_challenger(a,d,o), s(4):nil]
[e:combat(a,k,d,o,v), s(5):nil]
[f:reinitiate_trial(a,k,d,o,v), s(1):nil]
[f:reinitiate_trial(a,k,d,o,v), s(2):nil]
[s(1):nil, b:enter_worthy_defender(k,d,o)]
[s(1):nil, c:enter_beginner_defender(k,d,o)]
[s(2):nil, d:enter_challenger(a,d,o)]
[s(3):nil, e:combat(a,k,d,o,v)]
[s(4):nil, e:combat(a,k,d,o,v)]
[s(5):nil, f:reinitiate_trial(a,k,d,o,v)]
[s(5):nil, g:vindicate(d,o)]
[s(5):nil, h:condemn(d,o)]
[g:vindicate(d,o), end:nil]
[h:condemn(d,o), end:nil]
```

The process of dramatizing the Petri net representation of the plot involves a simple stepwise algorithm that controls the execution of the story by updating a list of active events according to a standard token-based execution approach. As described in Algorithm 1, function `Execute-PetriNet-Step` receives by parameter a Petri net `PN` and a list `C` with the nodes that were executed in the previous step of the algorithm (for the first step: `C = {start}`). The algorithm performs all the operations to activate place nodes and transition nodes for a single iteration of the execution process. The narrative events associated with activated transition nodes are added to set `A`, which is returned when the execution of the iteration ends. The set of narrative events returned by a single call of function `Execute-PetriNet-Step` represents the parallel events that take place during a certain point of the narrative. When the dramatization of these events ends, function `Execute-PetriNet-Step` can be called again in order to obtain the next narrative events for dramatization. If an empty set is returned, the narrative ends.

[4] https://love2d.org/

**Algorithm 1: Petri net execution algorithm**

```
function Execute-PetriNet-Step(PN, C)
  A = ∅;
  for each node V in C do
    if PN[V] is a PLACE then
      N = number of edges in PN[V];
      if N is greater than 0 then
        if N is 1 then
          S = first edge in PN[V];
        else
          S = get selected edge from PN[V] based on user interaction;
        end
        TA = get number of tokens available in parent nodes of PN[S];
        TN = get indegree of PN[S];
        if TA is greater or equal than TN then
          Consume TN tokens from the parent nodes of PN[S];
          Add S to A;
        end
      end
    else if PN[V] is a TRANSITION then
      for each edge E in PN[V] do
        if PN[E] is a PLACE then
          Add a token to place PN[E];
          L = Execute-PetriNet-Step(PN, {E});
          for each node W in L do
            Add W to A;
          end
        end
      end
    end
  end
  return A;
end
```

All the assets used for dramatization (e.g., character animations, background images, and audio files) are defined in a library manually constructed for the domain of a specific story. The *context library* is a 5-tuple $L = (\gamma, \alpha, \beta, \delta, \pi)$, where:

- $\gamma$ is a set that defines the *actors* of the story. Each actor has a name and a set of actions, which are represented by animations in a sprite sheet format;
- $\alpha$ defines the *locations* of the story. Besides associating each location with a background image and a soundtrack, it also defines a set of waypoints where actors can be placed during the scene composition process;
- $\beta$ defines the characters' dialogs (text and audio);
- $\delta$ is a set that defines the interaction points of the story. Each interaction point is associated with a set of interactive objects, which are represented by the classes of sketches that can be used by users to interact at each interaction point. The interaction points also include a set of instructions to guide the user during the interaction;
- $\pi$ establishes values for the variables present in the events of the Petri net.

In our implementation, the context library is defined in an XML file. The library used for the trial by combat example is available at: http://www.icad.puc-rio.br/~logtell/petri-net/context-trial-by-combat.xml

During the dramatization of the story, our system generates 2D animations in real time according to the actions performed by the virtual actors. An automatic virtual camera maintains the active actors always centered in the image frame while they move around the virtual world. When more than one actor is involved in the action, the camera will target the center of the scene, which is calculated based on the positions of all characters that are participating in the event. Figure 7 shows as an example a scene that involves two characters: Gawain and Guinevere.

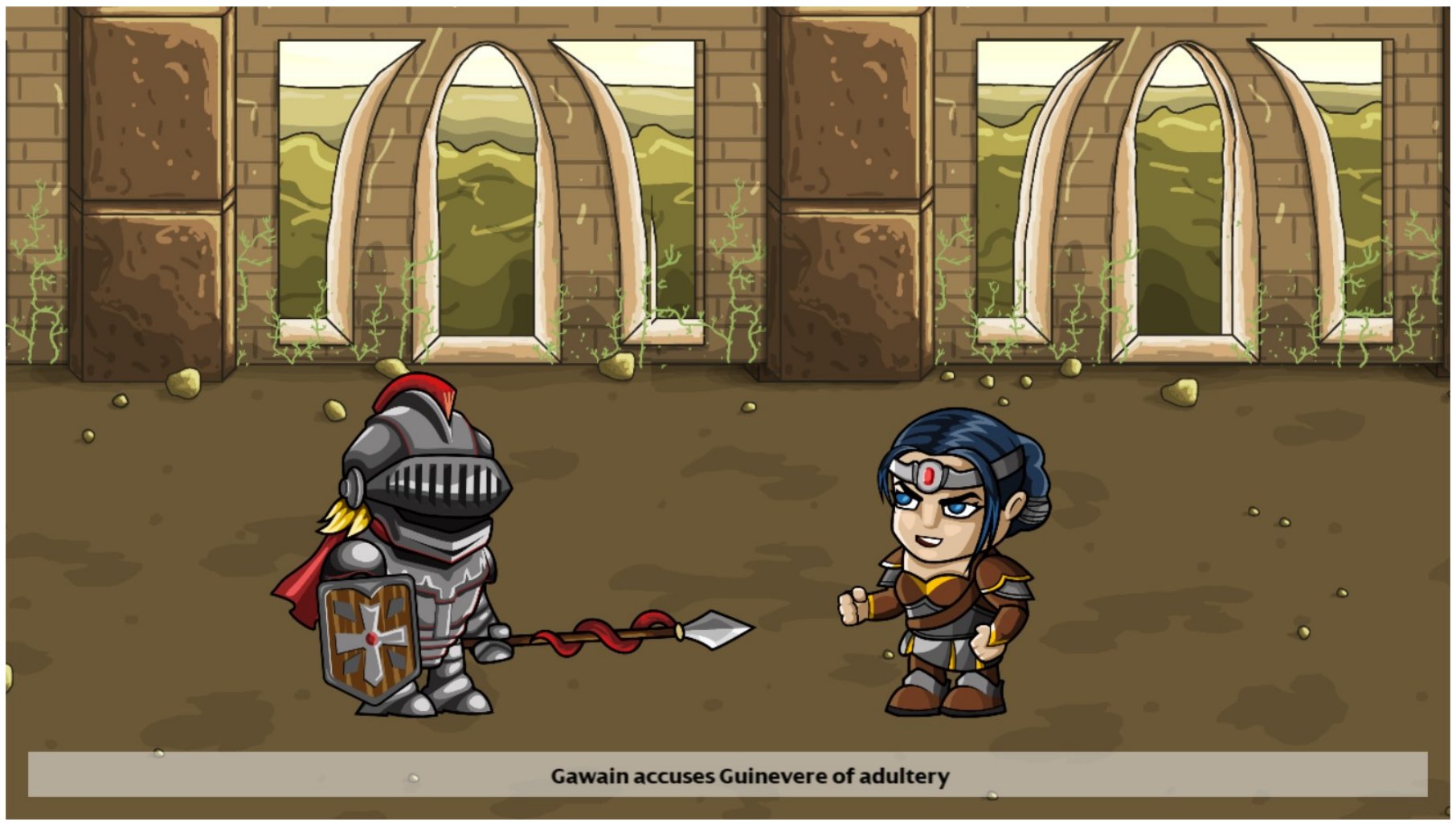

Figure 7: Scene from the dramatization of the trial by combat: Gawain accuses Guinevere of adultery

User interaction occurs at or-fork nodes of the Petri net. When a node of this type is activated, users are instructed by the virtual characters to interact by drawing specific objects in the interaction app. The instructions are defined in the context library and comprise a set of phases (text and audio), which are repeated until the user draws a valid object (single user interaction mode) or during a certain time frame (voting interaction mode). When the user's choice is identified, the corresponding transition node is selected to be activated (as indicated in line 10 of Algorithm 1). An example user interaction moment for the trial by combat is illustrated in Figure 8, showing the user's decision whether to help Gawain or Perceval (still in his squire's garb), in the combat by drawing a spear or a sword (i.e., the weapons used by each character). A full video demonstration of the trial by combat example is available at: https://www.youtube.com/watch?v=qI2TeBrhycc

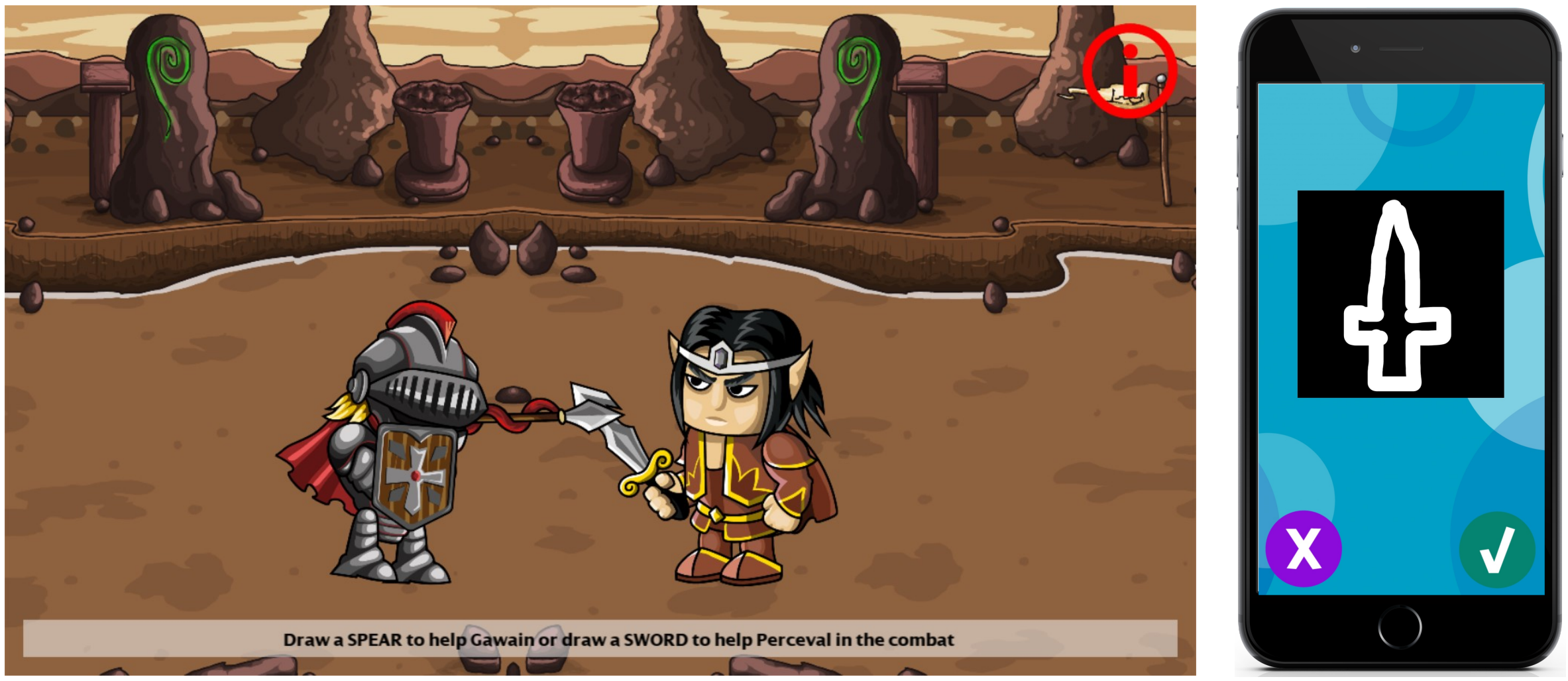

(*a*) (*b*)

Figure 8: User interaction moment in the trial by combat: (*a*) shows the dramatization system instructing the user to draw a spear to assist Gawain or a sword to help Perceval; and (*b*) shows that the user chose to draw a sword in the interaction app.

## 5. Concluding remarks

We claim that the research reported here provides a practically verifiable argument in favor of the complementarity between the situation calculus and the Petri net models. Indeed, the first can be used to generate the second.

The situation calculus approach is most convenient at specification time. The interplay of pre-conditions and post-conditions not only determines a semantically justified partial order of the defined actions, but also serves to enforce integrity constraints.

As a specified system just begins to be used, with a still empty log, process mining discovery would not be feasible. But, in lieu of the non-existent traces, generated plans could be used to derive a first Petri net model. Or, as done here, the Petri net can be directly derived from the situation calculus specification rules.

Plan generation has the necessary flexibility to investigate, before the system is liberated for official usage, whether the proposed specification allows all desirable use cases, and effectively disallows the illegal or, for some reason, undesirable ones. The price to pay is the time lost in exploring useless alternatives.

On the other hand, based on the conclusions of this investigation, Petri nets (like other workflow engines) can be designed to run in a tightly restrictive mode, with the additional asset of intuitive visualization of the workable sequences.

As an additional bonus whenever digital entertaining applications are treated, Petri nets are especially amenable to an attractive form of dramatization. In confirmation to this claim, a running system was developed, which served quite well to conduct experiments with the trial by combat narrative.

It must be recognized, however that, when discussing the Petri net model, we only dealt in this paper with the oversimplified trace representation used to introduce the most basic process mining notions at the beginning of Wil van der Aalst's paper. Since traces just register the name (or label) of the operations, therefore omitting the parameter lists, the value-dependent tests that are part of the interplay of preconditions and postconditions cannot be verified – in sharp contrast with the `check_fix` predicate (referred to at the end of section 2), programmed in terms of the situation calculus model.

As a proof of concept, we initiated this project working on the very limited version of the request processing example. Accordingly, we do not claim that all problems associated with more realistic applications, as well as more intricate Petri net schemes can be handled by the current prototype. For instance, Petri net loops caused by iterative actions have not been considered. It is expected that further research, covering more complex cases and extended representation conventions, will confirm the complementarity thesis and give evidence of its relevance to a fully satisfactory extent.

## Appendix

```
/* Simple interactive SWI-Prolog program for trace composition
   allowing to choose one of the two domains treated in the text:

   r: request processing
   t: trial by combat

   which share the same clausal Petri net representation */


:- set_prolog_flag(verbose,silent).
:- style_check([-singleton,-discontiguous]).
:-  set_prolog_flag(toplevel_print_options,[quoted(true),  portray(true),
max_depth(100), spacing(next_argument)]).
:- dynamic activated / 1 , enabled / 1, lop / 1 .

lop(init).

% special operators

:- op(900,fy,not).
:- op(650,yfx,=>).

% simplified reading utility

rl(L,L1) :-
  get0(A),
  (not (A == 10), !,
   append(L,[A],L2),
   rl(L2,L1);
   L1 = L).


% signatures of the operations

lop_request([register(c,   v,   t,   r),   examine_thoroughly(r,   c),
examine_casually(r, c), check_ticket(r, c, t),
     decide(r,   c,   v,   d),   reinitiate_request(r,   c,   t,   v),
pay_compensation(r, c, v), reject_request(r, c, v)]).

lop_trial([accuse(a,   d,   o),   enter_worthy_defender(k,   d,   o),
enter_beginner_defender(k, d, o), enter_challenger(a, d, o),
     combat(a, k, d, o, v), reinitiate_trial(a, k, d, o, v), vindicate(d,
o), condemn(d, o)]).


% Petri net clausal representation

i_arc(start, a).
arc(a, s(1), b).
arc(a, s(1), c).
arc(a, s(2), d).
arc(b, s(3), e).
arc(c, s(3), e).
arc(d, s(4), e).
arc(e, s(5), f).
arc(e, s(5), g).
arc(e, s(5), h).
```

```
arc(f, s(1), b).
arc(f, s(1), c).
arc(f, s(2), d).
e_arc(g, end).
e_arc(h, end).


% choosing the domain

:- nl,
   forall(clause(lop(X),Y,Z),erase(Z)),
   write('Please choose the domain:'),nl,
   write('  For request processing, type r'),nl,
   write('  For trial by combat, type t'),nl,
   write('my choice: '),
   rl([],N),
   name(Domain,N),
   (Domain = r, lop_request(L),!;
    Domain = t, lop_trial(L)),
  assert(lop(L)), !,
   nl,
   write('You can now compose one or more traces in the chosen domain'),
nl,
   write('each time entering:'), nl,nl,
   write('   traverse.'),nl,nl,!.


% Petri net interactive traversal

traverse :-
  traverse(T), !,
  nl,
  write(T),
  nl,
  trace_plan(T,P),
  nl,
  write(P), nl, nl, !.

traverse(X) :-
  nl,
  write(a),nl,
  check1(a),
  trav1(a,X),!,
  nl.

trav1(X,X) :-
  trav2(S),
  S == [],!.
trav1(X,Z) :-
  trav2(S),
  not (S = []),
  trav3(S,Es),
  length(Es,L),
  (L = 1,
   Es = [E],
   write(E),nl;
   L > 1,
   nl,write('choose one label from:'),nl,
   lab_op(Es,Esl),
```

```
    forall(member(Opl,Esl),
      (write(Opl),nl)),
    write('my choice: '),
    rl([],N),
    name(E,N)),
    atom_concat(X,E,Y),
    check1(Y),
    trav1(Y,Z).

trav2(S) :-
  enabled(E),
  e_arc(E,end),
  S = [], !.

trav2(S) :-
  setof(Se,
        Ei^(enabled(E),arc(E,Se,Ei)),
    Ss),
  (setof(Se,activated(Se),S1),!;
   S1 = []),
  append(Ss,S1,S).

trav3(S,Es) :-
  setof(E,
    (Ej,Si)^(arc(Ej,Si,E),
             member(Si,S)),
    Es1),
  setof(E,
    (Ej,Si)^(member(E,Es1),
              (forall(arc(Ej,Si,E),
                 member(Si,S)))),
   Es).

check1(T) :-
  (enabled(X), !,restart;
   not enabled(X)),
   i_arc(start,I),
   assert(enabled(I)),
   atom_chars(T,[I|ES]),
   ch1(ES).

ch1([]) :- !.
ch1([E|R]) :-
  activate(S),
  enable(S,E),
  ch1(R).

activate(S) :-
  setof(Se,
        Ei^(enabled(E),arc(E,Se,Ei)),
    Ss),
  forall(member(Se,Ss),
    assert(activated(Se))),
  setof(Se,activated(Se),S).

enable(S,Ei) :-
  retract(enabled(A)),
  forall(arc(Ej,Si,Ei),
         member(Si,S)),
```

```
  assert(enabled(Ei)),
  forall((member(Si,S),once(arc(Ej,Si,Ei))),
          retract(activated(Si))).
restart :-
  forall(clause(enabled(X),Y,Z),erase(Z)),
  forall(clause(activated(X),Y,Z),erase(Z)).

lab_op(L,Ls) :-
  lop(Ops),
  findall(Lab:Op,
     (member(Lab,L),
    name(Lab,[N]),
    N1 is N - 97,
    nth0(N1,Ops,Op1),
    Op1 =.. [Op|R]),
   Ls).


% converts trace notation into plan notation

trace_plan(T,P) :-
  lop(Ops),
  atom_chars(T,Lc),
  findall(O1,
   (member(C,Lc),
    name(C,[I1]),
   I is I1 - 97,
   nth0(I,Ops,O1)),
   L),
  plan_list(P,L).

plan_list(start,[]).
plan_list(Ts => T,V) :- !,
  plan_list(Ts,Vs),
  append(Vs,[T],V).
```